# Background-free fibre optic Brillouin probe for remote mapping of micromechanics

YuChen Xiang[1], Carin Basirun[2], Joshua Chou[2], Majid E. Warkiani[2], Peter Török[3,1], Yingying Wang[4], Shoufei Gao[4] and Irina V. Kabakova[5]

[1]Department of Physics, Blackett Laboratory, Imperial College London, Prince Consort Road, London SW7 2BW, UK

[2]School of Biomedical Engineering, Faculty of Engineering and IT, University of Technology Sydney, Ultimo, NSW 2007, Australia

[3]Nanyang Technological University, School of Physical and Mathematical Sciences, Singapore, 637371

[4]Institute of Photonics Technology, Jinan University, Guangzhou 510632, China

[5]School of Mathematical and Physical Sciences, University of Technology Sydney, Ultimo, NSW 2007, Australia

yuchen.xiang11@imperial.ac.uk
Carin.Basirun@uts.edu.au
Joshua.Chou@uts.edu.au
Majid.Warkiani@uts.edu.au
peter.torok@ntu.edu.sg
wangyy@jnu.edu.cn
gaosf@jnu.edu.cn

Corresponding author email: Irina.Kabakova@uts.edu.au
Corresponding author contact details: ph. +61426951911, building 4, office CB04.05.526, Harris Street, School of Mathematical and Physical Sciences, University of Technology Sydney, Ultimo, NSW 2007, Australia.

**Abstract:**
Brillouin spectroscopy is a century-old technique that has recently received renewed interest, as modern instrumentation has transformed it into a powerful contactless and label-free probe of micromechanical properties for biomedical applications. In particular, to fully harness the non-contact and non-destructive nature of Brillouin imaging, there is strong motivation to develop a fibre-integrated device and extend the technology into the domain of *in vivo* and *in situ* operation, such as for medical diagnostics. This work presents the first demonstration of a fibre optic Brillouin probe that is capable of mapping the mechanical properties of a tissue-mimicking phantom. This is achieved through combination of miniaturised optical design, advanced hollow-core fibre fabrication and high-resolution 3D printing. The protype probe is compact, background-free and possesses the highest collection efficiency to date, thus provides the foundation of a fibre-based Brillouin device for remote *in situ* measurements in challenging and otherwise difficult-to-reach environments, for biomedical, material science and industrial applications.

# Introduction

Brillouin spectroscopy has received much attention in the past decade as a novel non-contact and label-free method for mechanical probing of tissues, cells and biomaterials with cellular and subcellular resolution [1, 2]. Unlike contact methods such as elastography, micro-rheology and atomic force microscopy (AFM), Brillouin spectroscopy is completely non-destructive since it makes use of the light-induced acoustic fluctuations to retrieve the mechanical properties from the material. The information obtained is characteristic of relaxation behaviour of materials at high frequencies (GHz) and can be related to the material's compressibility and viscosity [3]. When combined with a standard confocal microscope, it is then possible to map these properties with optical resolution via a 3D scanning mechanism. The resultant Brillouin images are therefore hyperspectral and contain richer, more specific information owing to its mechanical contrast [4].

To date, Brillouin imaging has already been applied to a plethora of areas in life sciences, such as cells [5], human cornea [6], biofilms [7], zebra fish embryos [8], tumour spheroids [3], porcine cartilage [9] and assessment of atherosclerotic plaque formation in mouse arteries [10]. In the last case, a quantifiable difference between healthy and diseased arteries has been demonstrated on *ex vivo* samples, indicating that Brillouin imaging could guide cardiologists' medical assessment of the plaque's vulnerability and may assist in prediction of rupture likelihood. Specifically, instead of the conventional angiographic techniques, which produce low-resolution, non-specific images of coronary arteries [11], the development of a fibre-based diagnostic tool with enhanced imaging capabilities, such as the mechanical specificity and high spatial resolution, is envisioned to be a truly disruptive technology. In this vein, Brillouin imaging is highly attractive for being an all-optical modality, which allows for non-contact mechanical testing, at a distance. Undoubtedly, this property of Brillouin imaging is a strong advantage, compared to most other traditional elastographic methods [12]. Therefore, the possibility of Brillouin-based, *in vivo* imaging has continued to drive the development of a flexible, fibre optic probe not only useful for intravascular applications, but also for other scenarios that require remote and versatile operation (see a schematic of the fibre probe and the range of possible applications for remote micromechanical mapping in Figure 1).

Previously, we have demonstrated two proof-of-concept designs that is either bidirectional or uses separate fibres for light illumination and collection [13]. Although the design with a reciprocal light path is much simpler and more efficient by default, preliminary results have shown that the background signal in the fibre (generated by Brillouin and Raman spontaneous scattering) limits the achievable signal-to-noise-ratio (SNR) of the system. The dual-fibre design, although free from the fibre background, was found to be intrinsically inefficient due to the splitting of optical paths, resulting in large coupling loss and off-axis aberrations in the probe's focusing optics. More recently, a time-resolved, fibre-based Brillouin probe was proposed by La Cavera et al. [14], with which minute differences in the properties of liquid samples were detected. The probe functions, however, more akin to a fibre sensor as the probe tip was required to contact the sample to facilitate local heating, in order to generate the signal.

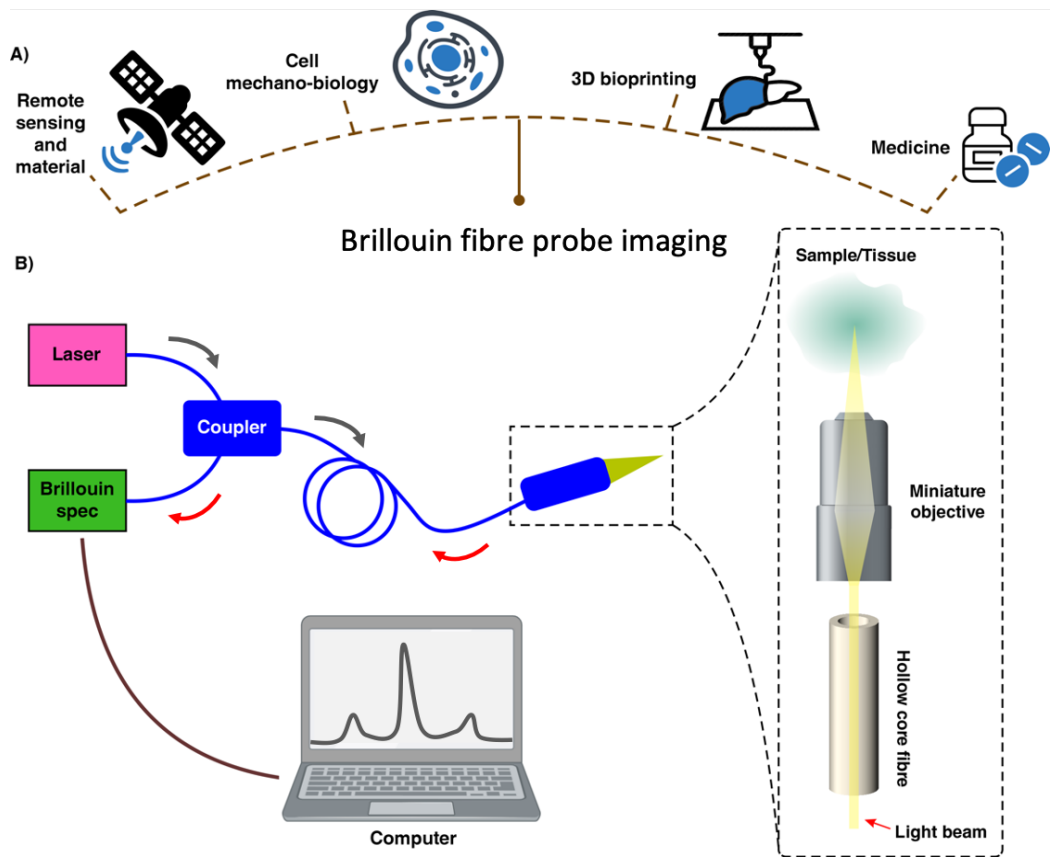

**Figure 1. Brillouin fibre probe design and potential applications.** *A) Brillouin fibre probe can be integrated with existing technologies for remote sensing and biomaterial manufacturing or serve as a novel solution for a wide range of applications in biomedicine [6, 9, 10]. B) Schematic of an imaging system based on Brillouin fibre probe. The inset shows a hollow-core fibre and a system of miniaturised lenses to enable remote, contact-free measurements.*

Here we present the first demonstration of a Brillouin fibre probe capable of micro-mechanical scanning of biological phantoms and that utilises a new conceptual design based on a hollow core fibre (HCF) (see Figure 1 for details of the design). The HCF is a result of the advancement in fibre technology, especially in terms of fabrication, as this type of waveguide can confine light without the conventional total internal reflection guidance, but rather by utilising photonic bandgap or anti-resonant guidance in the central low refractive index hollow-core region surrounded by glass photonic crystal structures [15] or anti-resonant elements in the cladding [16]. Since the core region of the HCF consists only of air, the problem of the fibre scattering background in a bidirectional Brillouin probe is inherently remedied. In addition, we incorporate a new nodeless hollow-core fibre design [17] that has achieved sufficiently low propagation and bending losses that helps to achieve high collection efficiency in the Brillouin fibre imaging system, making the throughput comparable to the free-space optical Brillouin microscopy setups [18]. We believe, therefore, that the hollow-core fibre optical Brillouin probe presented in this study is the first key milestone in the new area of flexible and miniature Brillouin imaging systems, suitable for the wide range of medical, industrial and bioengineering applications where real-time, non-destructive, *in situ* micro-mechanical characterisation is required (see Figure 1 (A) for potential applications).

# Results

Firstly, we characterise the spectral background of the HCF system to validate that it does not pose a problem for Brillouin microspectroscopy. To demonstrate this, multiple spectra were taken without any sample at the distal side with an incident power of 20mW from the laser, of which approximately 12mW was successfully coupled into the HCF, yielding an input coupling efficiency of ($\eta = 60\%$). The signal was acquired for over 2 minutes to achieve sufficient SNR for fitting and the results can be found in Figure 2A, represented by the blue solid curve. Most notably, the problematic, wideband background that is characteristic of Raman scattering in silica fibres [19] can no longer be observed. There are, however, some weak residual spectral features at two distinct locations within the free spectral range (FSR) of the spectrometer (from -37.78 GHz to +37.78 GHz). As the Brillouin doublet is symmetrically positioned either side of the laser frequency, only the positively shifted peaks (Anti-Stokes) are presented in the figures for conciseness. Namely, there are identifiable Brillouin peaks appearing at 10.82 GHz and further away at around 25 GHz. Upon closer inspection the latter consists of two closely spaced features at 23.93 and 25.74 GHz, respectively. Qualitatively, the more prominent features towards the edges of the FSR correspond to relatively fast phonon modes and appear to be characteristic of common glass materials [20], whereas the other background signal resembles that of plastic materials (e.g. poly-acrylate plastics) previously measured in our lab. It is then reasonable to assume that these features correspond to the Brillouin scattering in the glass anti-resonant structure that serves as an effective cladding in the fibre and in the outer, protective polymer jacket respectively.

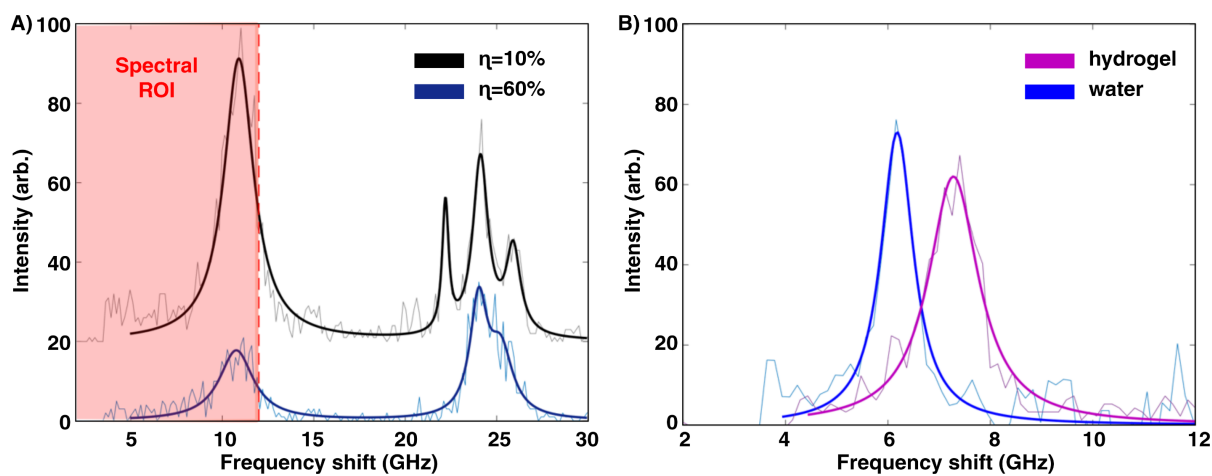

**Figure 2: Brillouin spectra acquired using the hollow-core fibre optic probe.** *A) Background spectra at different coupling efficiencies ($\eta$); B) Spectra of water and GelMA hydrogel acquired after background removal.*

To test this hypothesis, the optics for coupling into the fibre was deliberately defocused to decrease the coupling efficiency to $\eta = 10\%$, which led to a much larger amount of light in the cladding and outer jacket regions. The resultant spectrum acquired over the same amount of time is also shown in the same figure with an arbitrary offset for clarity (black solid curve), where a visible increase in intensity can be observed for all background features as expected, albeit not proportionally, due to the different propagation lengths of cladding and jacket modes in the HCF, which makes the estimation of effective scattering volumes difficult. Interestingly, apart from confirming the existence of at least two modes in the far region, the background spectrum also sees the introduction of an additional mode at 21.90 GHz. These different glass-induced Brillouin peaks can be attributed to the existence of higher order modes propagating in the fibre. Existence of such modes have been reported previously for short sections of the fibre with a similar design and can be attributed to incomplete fulfilment of the anti-resonance condition due to deviation of the fibre fabrication from the design parameters [21]. These higher order modes ($LP_{11}$, $LP_{02}$, $LP_{20}$) differ from the fundamental mode in terms of propagation constants and thus can be phase-matched to different phonon modes according to a continuous dispersion relation in the same glass material. All of the glass modes, however, do not present any significant hurdle to Brillouin microspectroscopy and imaging, since they are located sufficiently far away from the spectral region of interest (ROI=12 GHz) where peaks for common soft materials are expected.

The mode associated with the polymer jacket of the HCF, however, presents inconvenience for fibre-integrated Brillouin imaging as its spectral region overlaps with the ROI. To reduce the background associated with the fibre jacket to a negligible level, the first 10 cm of the HCF fibre at the proximal side of the fibre were stripped, which was sufficient to almost fully eradicate the plastic mode due to its short propagation lengths in the fibre. Additional spatial filtering was also implemented before the entrance aperture of the spectrometer, which induces a further 30% decrease in systematic throughput. Any non-background signals, however, are largely unaffected, as only the non-fundamental modes possess different numerical apertures (NAs) and generally traverse on the outer rim of the beam. Any remaining background was easily filtered by software during post-processing as the contaminants are known.

To calibrate the spectral response of the system, the signal from distilled water and hydrogel were recorded by the probe with 8mW at the sample, and an acquisition time of 2 minutes to facilitate accurate line fitting. Both spectra are presented in Figure 2B. Spectral analysis of the measurement data for water produces an average frequency shift value of 6.01 GHz with a linewidth of 0.733 GHz. While the spectral shift agrees well with the value expected for the laser wavelength (6.05 GHz at 660 nm) at room temperature [22], the width of the peak is wider than the theoretical value due to instrumental broadening caused by the imaging system and the spectrometer optics [23]. As the linewidth can be linked to the viscosity of the liquid [24] and provides extra information from the sample, the known width of water can be used to deconvolve and retrieve the instrumental response function, assuming that the system is linear [25]. By performing deconvolution as described in the Supplementary materials, the response function was approximated to follow a Lorentzian distribution with a linewidth of 0.626 GHz in the spectral ROI. To validate this calibration procedure, the Brillouin parameters obtained from gelatin-based hydrogel (GelMA) are compared. The Brillouin shift and linewidth of the sample were determined to be 6.97 GHz and 1.25 GHz respectively. Taking into account the spectral response of the system, the true linewidth of the hydrogel is then 0.624 GHz, which agrees with the value calculated from the viscoelastic parameters of the GelMA hydrogel [26].

To demonstrate the ability of the probe to obtain spatial information as well as spectral information, surface profilometry of a hydrogel-water phantom was performed by mounting the sample on a translation stage. The hydrogel-water phantom has been created to mimic the mechanical properties of biological tissues, since both possess high water content and a similar range of mechanical responses at GHz frequencies. A sketch of the phantom morphology is shown in Figure 3A and both a coarse scan and a fine scan were performed along the directions indicated in the schematics, with step sizes of 0.5 mm and 0.2 mm respectively. 10 mW of incident power was focused onto the mounted slide and the sample was brought to focus such that maximum signal strength was achieved. The starting position was identified visually so that the line scan would go through the centre of the hydrogel droplet. The acquisition time for each spectrum was chosen to achieve a consistent SNR (~25). As the gel region produces a slightly weaker signal due to higher turbidity, the acquisition time thus varied between 30-40 seconds per point. The surface profile of the droplet can be visualised by simply displaying the raw spectra obtained along the direction of scanning, as heat maps in Figure 3B.

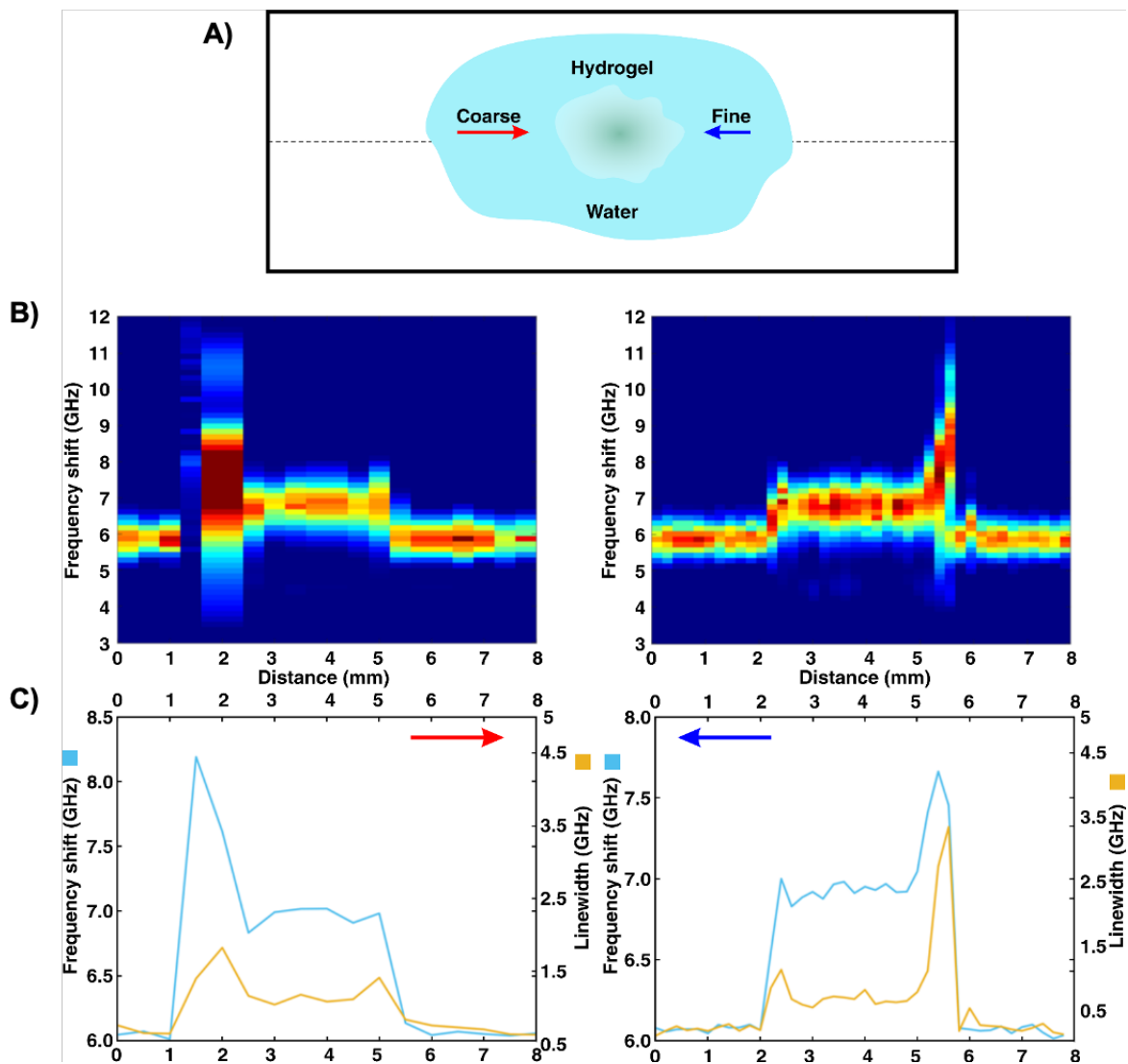

**Figure 3: Surface profilometry results of the hydrogel-water phantom.** *A) Illustration of the sample orientation, the axis of scanning (dotted line) and scanning directions (red and blue arrows for course and fine scans, respectively). B) Spectral heat maps. C) Brillouin shift & linewidth profiles.*

Figure 3b are such heat maps showing the Anti-Stokes peaks of the spectra that are generated by first denoising the spectral ROI [27]. It can be seen that there is a clear contrast that identifies the hydrogel region even with the coarser scan. Interestingly, the seemingly spurious features at the edges of the droplet are reproducible and can be explained by more solid layer of gel produced at the edges of the droplet during UV crosslinking stage (see methods for sample preparation). Quantitatively, it is also possible to plot the surface profile of the phantom in terms of the Brillouin shift and linewidth values, both are presented below in Figure 3C, by blue and yellow lines respectively. Apart from recreating the profile shape observed in the raw data, from the values obtained with the fine scan, the water region gives an average shift of 6.08 GHz, with a standard deviation of 0.02 GHz. In the hydrogel region, excluding the outliers at the edges, an average shift of 6.92 GHz and standard deviation of 0.04 GHz are calculated. In terms of the spectral widths, the fitting yields average linewidths of 0.0835 GHz and 0.489 GHz, with standard deviations of 0.0466 GHz and 0.067 GHz for water and hydrogel respectively, after removing the measurement system response function.

## Discussion

To the best of our knowledge, we have demonstrated, for the first time, a fibre optic Brillouin probe that can be used to simultaneously collect spectral and spatial information in a non-contact manner. The spectral performance of the system is directly comparable to that of a free-space system and no longer suffers from the background signal generated by spontaneous Brillouin and Raman scattering in the fibre that was detrimental in previous designs [13]. The probe is sensitive to both changes in the Brillouin shift and linewidth, which all can contain information about the viscoelastic properties of the sample. The spatial sensitivity is facilitated by a custom-designed, miniaturised objective (see Supplementary material), which retains a sufficient working distance for non-contact signal collection while producing a focal spot just over 10μm, which governs the spatial resolution of the probe. A straightforward improvement could be to modify the optical design in order to achieve higher resolution for imaging applications. Such designs are commercially available and able to achieve an NA as high as 0.3 [28], which would make the resolution of the miniaturised system more comparable to the microscope equivalent. For some samples, such as arterial vessels, rather than higher resolution, it is usually more attractive to have a high acquisition speed for real-time operation. To improve on this particular aspect in future implementations, both the system throughput and the scanning mechanisms have to be enhanced. For instance, a VIPA (Virtually-Imaged Phase Array) based spectrometer can be incorporated which would enable acquisition speed on the order of tens of milliseconds per spectrum [18]. A fast, resonant fibre scanning variant also appears to be more favourable as they have already been demonstrated to function well with a confocal fibre system while maintaining the flexibility and dimension of a single fibre [29]. Finally, the current design can also be converted into a multimodal instrument. Fibre-based Raman probes have been developed since the 1990s and thus benefit from being a more mature technology [30]. The inclusion of a Raman channel, for example by adding a ring of collection fibres around the HCF, would be an ideal addition to our probe. The chemical sensitivity of the Raman mode would allow the quantitative monitoring of water content in the sample, which has been demonstrated to be a dominant factor in the Brillouin shift values measured [31] and may help in the retrieval of pure mechanical properties in the future.

## Methods and materials

**Fibre-based Brillouin system**

As illustrated in Figure 4, the laser (Torus 660 nm, Laser Quantum) produces a collimated beam with a diameter of 1.66 mm and maximum power of 120 mW. A linear polariser (LP) is placed directly after the laser input to ensure a linear polarisation state, which is crucial to facilitate polarisation control in the system later on. A non-polarising beam-splitter (BS) with a split ratio of 90:10, for transmission and reflection respectively, then directs light into the spectrometer (TFP1) for alignment. The transmitted beam is then expanded to twice of its diameter with the two lenses described in the design section. A polarising beam splitter (PBS), whose transmission axis is aligned with the orientation of the LP, is expected to transmit 99.9% of the incident light. A quarter waveplate ($\lambda/4$) placed at 45 degrees with respect to LP then converts the incident linear polarisation state to circularly polarised state. A single achromatic doublet (f=75mm, AC254-075-A-ML, Thorlabs) is then used to couple light to the hollow core fibre which was in-house fabricated to produce a mode field diameter (MFD) of ~20 µm and NA~0.02 at 660nm. The properties and performance of the fibre are comparable to that described in a previous publication [32]. The fibre structure and beam profile after the fibre are also consistent with the measurements reported previously in [32] (Figure 4C).

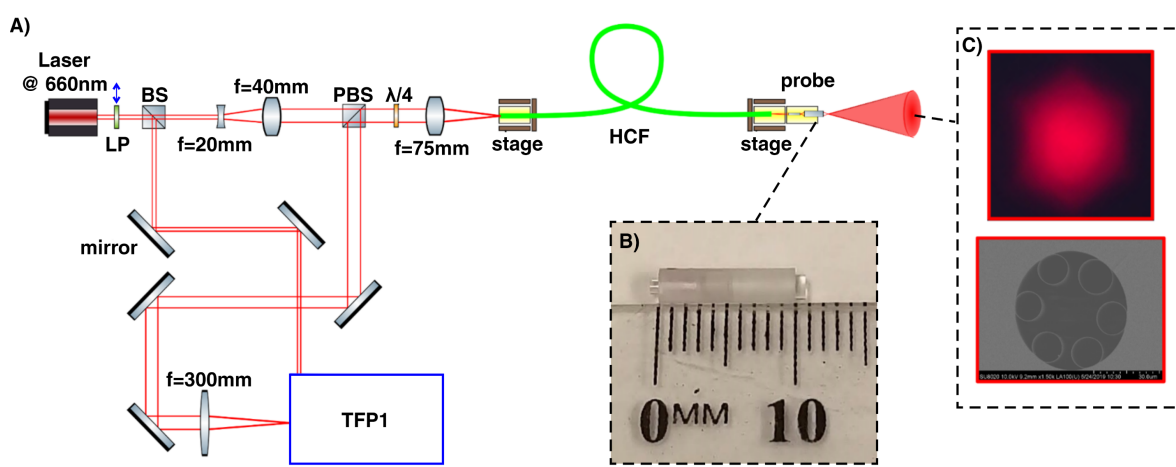

**Figure 4: Experimental configuration of the fibre-based Brillouin set-up.** *A) The optical set-up. B) The distal probe assembly. C) The beam profile & SEM image of the HCF.*

The entire length of approximately 3 metres were used initially to ensure single mode operation, the fibre was stripped and cleaved at 90° at both ends and carefully mounted onto two 3-axis flexure stages (MAX313D, Thorlabs) for fine alignment. The input fibre coupling efficiency was measured to be ≥60%. Lastly, the distal probe assembly consists of two GRIN lenses (G2P10, GRIN2906, Thorlabs) inserted into the transparent, 3D printed sleeve (Microfluidics BV-003). A photograph of the assembled probe is presented in Figure 4B. The same optics serves both to illuminate and collect back-scattered light from the sample, which is back-coupled into the HCF (collection efficiency ~70%) and sent back along the same light path. Once the scattered light arrives at the waveplate, it is rotated back into a linear polarisation state, but at a direction orthogonal to LP, which causes the PBS to direct the back-scattered light towards the spectrometer arm, where the large doublet lens (f=300mm, AC508-300-AB-ML, Thorlabs) finally focuses it into the spectrometer for analysis.

**Optical design aspects**
The optical design aspect of our instrument is best discussed by dividing the set-up into two halves, the proximal side and the distal side, with the latter being the sample facing direction. The optics at the proximal side are designed for the sole purpose of efficiently coupling light into the spectrometer and the fibre. The spectrometer used in this work is a tandem Fabry-Perot (FP) interferometer (TFP-1) supplied by JRS Scientific Instruments. The system is equipped with an adjustable entrance aperture, ranging from 0.07 mm to 1.3 mm in size. Just as for a grating spectrometer, in a FP system, both the spectral resolution and the throughput are largely determined by the shape and size of the input aperture [33]. In the interest of maximising throughput and hence SNR of the device, the largest aperture size (1.3 mm) is to be used for all experiments and the spectral response of the system can be characterised by acquiring the spectrum of a known substance, namely distilled water at room temperature. A commercial achromatic (AC) doublet (AC508-300-AB-ML, Thorlabs) with f=300mm is thus chosen to focus onto the entrance pinhole, due to the large tolerance on the depth of focus which assists in system alignment. Similarly, the appropriate optics is needed to match the MFD of the HCF [32]. Another system is then designed to first expand the beam from the laser (diameter=1.66 mm) by a factor of 2 and then it is focused into the fibre with a f=75mm AC doublet (AC254-075-A-ML, Thorlabs) to achieve a theoretical coupling efficiency of 82.2%.

While standard optical components can be used at the proximal side, the distal optics requires the use of miniaturised optics to achieve the dimension that is necessary for any endoscopic devices. In previous work, we demonstrated that the use of a single graded-index (GRIN) lens after a single mode fibre was sufficient to facilitate the illumination and collection for the purpose of spectroscopy. This simple arrangement, however, suffers from large off-axis aberrations and is not directly applicable to the HCF-based system, due to the unusually low NA of our HCF fibre. Therefore, the distal optics for the new device was re-designed with the following considerations in mind: 1) high collection efficiency; 2) image space NA≥0.1; 3) probe working distance ≥1 mm; 4) probe diameter < 2 mm and 5) adequate tolerances for alignment and fabrication of the probe housing (3D printed sleeve).

To deal with the low NA of the fibre source, a design with two GRIN lenses with different NAs was adopted to achieve the image space NA while retaining the working distance, which functions as a simple miniaturised objective. Two off-the-shelf lenses with different profiles (G2P10 and GRIN2906, Thorlabs) were chosen for this design, which was then simulated using *Zemax OpticStudio.* To compute the collection efficiency of the design, a perfect mirror is placed at the sample plane, which effectively causes the rays to perform a double-pass in the system that should be taken into account when considering transmission losses. With an equal weighting applied to achieve an image space NA of 0.1 and maximum back-coupling efficiency, the resultant optimised optical set-up can be found in the Supplementary materials.

For the mini-objective lens it was found that there is a trade-off between imaging NA and aberrations as the fibre source is moved closer to the first lens. Therefore, the optimal distance that minimises aberrations while retaining the focusing ability was found to be 5.5mm, with the second lens should be placed 2.25mm away from the first. At the sample side, this translates to a diffraction limited spot radius of 5.081µm, with NA=0.08 and working distance of 1.04 mm. This focal plane is then conjugated with the entrance of the HCF, which acts as a pinhole with a diameter of ~20µm. The back-coupling efficiency is maximised in this arrangement to be 87.0%. Since the first lens is antireflection coated on both sides, the only other loss in transmission comes from the two surfaces of the second lens, which is expected to sum to about 7% for the incident wavelength. In summary, as well as achieving a higher resolution than before, the maximum collection efficiency of this probe is expected to be 0.87x0.93=81.0%, which is an order of magnitude higher than what was achieved with our dual-fibre design in the past. The tolerance of the design was verified by inducing decenter & axial misalignment of ≤0.5mm and relative tilt up to 1° on the elements in *Zemax OpticStudio.* By performing Monte Carlo simulations of 20 different configurations within these ranges, the focal spot was found to be still diffraction-limited on average, while the worst-case scenario produces an RMS radius of 6.98 µm at the sample plane. In terms of the back-coupling efficiency, the expected value with these large tolerances only experiences a decrease of 8% when compared to the ideal case, which is still sufficiently high. To fix the relative alignment of the distal optics, a custom-designed sleeve structure was 3D printed in-house using high precision stereolithography according to the design presented in the Supplementary materials. The tolerance of the fabrication process allows for an alignment accuracy of $\pm 15 \mu m$, which is well within the predicted tolerances.

**Sample preparation**
The spectroscopy samples used in this study are either pure, distilled water or hydrogels prepared in-house. The former can be easily deposited in the same V-groove mount that holds the probe assembly on the flexure stage, where surface tension alone is sufficient to hold the droplet in place. The latter is the GelMA (0.25% LAP) hydrogel, printed using a BioX bioprinter from Cellink, using the droplet protocol available. We optimised the best printing parameters to be at 28°C and a pressure of 30kPa. The extrusion time was adjusted to be 1 second to account for the sample size. The droplet sample was crosslinked by exposure to UV light (405 nm UV module) available on the BioX and it was exposed for approximately 10 minutes at a distance of 15 cm from the sample. The final mixture produced a hydrogel with ~10% in solid fraction.

For spatial scanning, a combination of the two materials was used to create a `phantom' with spatially contrasting mechanical properties. A droplet of the GelMA hydrogel, roughly 5mm in diameter was first prepared on a microscopy slide as described above. Next, a multilayer tape with a square window for the droplet was deposited on the slide. A small amount of distilled water was dropped with a pipette within the window area to immerse and hydrate the hydrogel droplet. A 0.14 mm microscopy cover slip was finally placed on top to seal the structure and prevent water evaporation. We also took special care to remove any air bubbles within water-hydrogel construct.

**Data analysis**
Data analysis is performed in *Matlab* using a custom-written program. The spectrometer commercial software (GHOST) outputs data which are already frequency-calibrated. These raw spectra are first pre-processed, specifically background subtracted and denoised using wavelet analysis according to protocols presented in [27], and then spectrally analysed using least-squares line fitting [34].

## Statement of author contributions
YCX, PT and IVK conceived the idea of this work. YCX and IVK performed the experiments and written the manuscript. YCX designed optical systems for the HCF coupling and for the miniature fibre probe, created software for data analysis and performed all signal processing, spectral fitting and denoising routines. CB prepared hydrogel samples. MEW fabricated the packaging for the miniature fibre probe by high-resolution 3D printing and contributed to creation of illustrations. YW and SG fabricated the hollow-core fibre. All authors contributed to the results discussion and editing of the manuscript.

## Acknowledgments
This work was supported by the Australian Research Council Discovery Program (DP190101973) and the University of Technology Sydney, Faculty of Science 2019 Seed Award. In addition, we express special gratitude to Dr Eric Magi and Dr Moritz Merklein for their valuable advices on the speciality fibre handling and for the use of fibre optic equipment.

## Conflict of Interest
The authors declare that they have no conflict of interest.

## References

[1]  Z. Meng, A. J. Traverso, C. W. Ballmann, M. A. Troyanova-Wood and V. V. Yakovlev, "Seeing cells in a new light: a renaissance of Brillouin spectroscopy," *Advances in Optics and Photonics,* vol. 8, no. 2, p. 300, 2016.

[2]  F. Palombo and D. Fioretto, "Brillouin Light Scattering: Applications in Biomedical Sciences," *Chemical Reviews,* vol. 119, no. 13, pp. 7833-7847, 10 7 2019.

[3]  J. Margueritat, A. Virgone-Carlotta, S. Monnier, H. Delanoë-Ayari, H. C. Mertani, A. Berthelot, Q. Martinet, X. Dagany, C. Rivière, J. P. Rieu and T. Dehoux, "High-Frequency Mechanical Properties of Tumors Measured by Brillouin Light Scattering," *Physical Review Letters,* vol. 122, no. 1, pp. 1-6, 2019.

[4]  K. J. Koski and J. L. Yarger, "Brillouin imaging," *Applied Physics Letters,* vol. 87, no. 6, pp. 1-4, 2005.

[5]  G. Scarcelli, W. J. Polacheck, H. T. Nia, K. Patel, A. J. Grodzinsky, R. D. Kamm and S. H. Yun, "Noncontact three-dimensional mapping of intracellular hydromechanical properties by Brillouin microscopy," *Nature Methods,* vol. 12, no. 12, pp. 1132-1134, 2015.

[6]  R. M. Gouveia, G. Lepert, S. Gupta, R. R. Mohan, C. Paterson and C. J. Connon, "Assessment of corneal substrate biomechanics and its effect on epithelial stem cell maintenance and differentiation," *Nature Communications,* vol. 10, no. 1, p. 1496, 3 12 2019.

[7]  A. Karampatzakis, C. Z. Song, L. P. Allsopp, A. Filloux, S. A. Rice, Y. Cohen, T. Wohland and P. Török, "Probing the internal micromechanical properties of Pseudomonas aeruginosa biofilms by Brillouin imaging," *npj Biofilms and Microbiomes,* 2017.

[8] R. Raghunathan, J. Zhang, C. Wu, J. Rippy and M. Singh, "Evaluating biomechanical properties of murine embryos using Brillouin microscopy and optical coherence tomography," *Journal of Biomedical Optics,* vol. 22, no. 08, p. 1, 2017.

[9] P.-J. Wu, M. I. Masouleh, D. Dini, C. Paterson, P. Török, D. R. Overby and I. V. Kabakova, "Detection of proteoglycan loss from articular cartilage using Brillouin microscopy, with applications to osteoarthritis," *Biomedical Optics Express,* vol. 10, no. 5, p. 2457, 1 5 2019.

[10] G. Antonacci, R. M. Pedrigi, A. Kondiboyina, V. V. Mehta, R. de Silva, C. Paterson, R. Krams and P. Török, "Quantification of plaque stiffness by Brillouin microscopy in experimental thin cap fibroatheroma," *Journal of The Royal Society Interface,* vol. 12, no. 112, p. 20150843, 6 11 2015.

[11] O. Ghekiere, R. Salgado, N. Buls, T. Leiner, I. Mancini, P. Vanhoenacker, P. Dendale and A. Nchimi, "Image quality in coronary CT angiography: challenges and technical solutions," *The British Journal of Radiology,* vol. 90, no. 1072, p. 20160567, 4 2017.

[12] B. Kennedy, K. Kennedy, D. Sampson, "A review of optical coherence elastography: Fundamentals, techniques and prospects", IEEE Journal of Selected Topics in Quantum Electronics, vol. 20, p.1-17, 2014.

[13] I. V. Kabakova, Y. Xiang, C. Paterson and P. Török, "Fiber-integrated Brillouin microspectroscopy: Towards Brillouin endoscopy," *Journal of Innovative Optical Health Sciences,* vol. 10, no. 06, p. 1742002, 2017.

[14] S. La Cavera, F. Pérez-Cota, R. Fuentes-Domínguez, R. J. Smith and M. Clark, "Time resolved Brillouin fiber-spectrometer," *Optics Express,* vol. 27, no. 18, p. 25064, 2019.

[15] P. Russell, T. A. Birks and J. C. Knight, "Photonic crystal fibres," *Nature,* vol. 424, no. August, pp. 847-851, 2005.

[16] W. Ding, Y.-Y. Wang, S.-F. Gao, M.-L. Wang and P. Wang, "Recent Progress in Low-Loss Hollow-Core Anti-Resonant Fibers and Their Applications," *IEEE Journal of Selected Topics in Quantum Electronics,* vol. 26, no. 4, pp. 1-12, 7 2020.

[17] S.-F. Gao, Y.-Y. Wang, X.-L. Liu, W. Ding and P. Wang, "Bending loss characterization in nodeless hollow-core anti-resonant fiber," *Optics Express,* vol. 24, no. 13, p. 14801, 27 6 2016.

[18] Z. Coker, M. Troyanova-Wood, A. J. Traverso, T. Yakupov, Z. N. Utegulov and V. V. Yakovlev, "Assessing performance of modern Brillouin spectrometers," *Optics Express,* vol. 26, no. 3, p. 2400, 5 2 2018.

[19] R. H. Stolen, W. J. Tomlinson, H. A. Haus and J. P. Gordon, "Raman response function of silica-core fibers," *Journal of the Optical Society of America B,* vol. 6, no. 6, p. 1159, 1989.

[20] D. Heiman, D. S. Hamilton and R. W. Hellwarth, "Brillouin scattering measurements on optical glasses.," *Physical Review B: Condensed Matter and Materials Physics,* vol. 19, no. 12, pp. 6583-6592, 1979.

[21] A. Van Newkirk, J. E. Antonio-Lopez, J. Anderson, R. Alvarez-Aguirre, Z. S. Eznaveh, R. Amezcua-Correa and A. Schulzgen, "Modal analysis of anti-resonant hollow core fibers," *2016 Conference on Lasers and Electro-Optics, CLEO 2016,* vol. 41, no. 14, pp. 3277-3280, 2016.

[22] K. J. Koski, "Brillouin Scattering Database," [Online]. Available: http://koski.ucdavis.edu/BRILLOUIN/water/water.html

[23] P. A. Wilksch, "Instrument function of the Fabry-Perot spectrometer," *Applied Optics,* vol. 24, no. 10, p. 1502, 15 5 1985.

[24] R. D. Mountain, "Thermal relaxation and brillouin scattering in liquids," *Journal of Research of the National Bureau of Standards Section A: Physics and Chemistry,* 1966.

[25] P. Török and M. R. Foreman, "Precision and informational limits in inelastic optical spectroscopy," *Scientific Reports,* vol. 9, no. 1, pp. 1-16, 2019.

[26] S. M. Naseer, A. Manbachi, M. Samandari, P. Walch, Y. Gao, Y. S. Zhang, F. Davoudi, W. Wang, K. Abrinia, J. M. Cooper, A. Khademhosseini and S. R. Shin, "Surface acoustic waves induced micropatterning of cells in gelatin methacryloyl (GelMA) hydrogels," *Biofabrication,* vol. 9, no. 1, p. 015020, 14 2 2017.

[27] Y. Xiang, M. R. Foreman and P. Török, "SNR enhancement in brillouin microspectroscopy using spectrum reconstruction," *Biomedical Optics Express,* vol. 11, no. 2, p. 1020, 1 2 2020.

[28] G. Matz, B. Messerschmidt and H. Gross, "Design and evaluation of new color-corrected rigid endomicroscopic high NA GRIN-objectives with a sub-micron resolution and large field of view," *Optics Express,* vol. 24, no. 10, p. 10987, 2016.

[29] C. M. Lee, C. J. Engelbrecht, T. D. Soper, F. Helmchen and E. J. Seibel, *Scanning fiber endoscopy with highly flexible, 1 mm catheterscopes for wide-field, full-color imaging,* vol. 3, 2010, pp. 385-407.

[30] I. Latka, S. Dochow, C. Krafft, B. Dietzek and J. Popp, "Fiber optic probes for linear and nonlinear Raman applications - Current trends and future development," *Laser and Photonics Reviews,* vol. 7, no. 5, pp. 698-731, 2013.

[31] P.-J. Wu, I. V. Kabakova, J. W. Ruberti, J. M. Sherwood, I. E. Dunlop, C. Paterson, P. Török and D. R. Overby, "Water content, not stiffness, dominates Brillouin spectroscopy measurements in hydrated materials," *Nature Methods,* vol. 15, no. 8, pp. 561-562, 31 8 2018.

[32] S.-f. Gao, Y.-y. Wang, X.-l. Liu, C. Hong, S. Gu and P. Wang, "Nodeless hollow-core fiber for the visible spectral range," *Optics Letters,* vol. 42, no. 1, p. 61, 2017.

[33] I. Aramburu, M. Lujua, G. Madariaga, M. A. Illarramendi and J. Zubia, "Effect of finite beam size on the spatial and spectral response of a Fabry-Perot interferometer," in *14th Conference on Education and Training in Optics and Photonics: ETOP 2017*, 2017.

[34] T. O'Haver, Pragmatic Introduction to Signal Processing, Retirement project, 2019.